\documentclass[10pt,english,prl,showpacs,superscriptaddress]{revtex4}
\usepackage[latin9]{inputenc}
\usepackage{textcomp}
\usepackage{amsmath}
\usepackage{amssymb}
\usepackage{graphicx}

\makeatletter
\@ifundefined{textcolor}{}
{%
 \definecolor{BLACK}{gray}{0}
 \definecolor{WHITE}{gray}{1}
 \definecolor{RED}{rgb}{1,0,0}
 \definecolor{GREEN}{rgb}{0,1,0}
 \definecolor{BLUE}{rgb}{0,0,1}
 \definecolor{CYAN}{cmyk}{1,0,0,0}
 \definecolor{MAGENTA}{cmyk}{0,1,0,0}
 \definecolor{YELLOW}{cmyk}{0,0,1,0}
}

\usepackage{epstopdf}
\usepackage{color}
\makeatother

\usepackage{babel}
\begin{document}

\title{Plant clonal morphologies and spatial patterns as self-organized responses to resource-limited environments}

\author{
P. Couteron$^{1}$, F. Anthelme$^{1,2}$,  M. Clerc$^{3}$, D. Escaff$^{4}$, C. Fernandez-Oto$^{5}$ and M. Tlidi$^{5}$}

\address{$^{1}$IRD - UMR AMAP, c/o Cirad TA A-51/PS2, Boulevard de la Lironde, 34398 Montpellier cedex, France.\\
$^{2}$Universidad Major de San Andr{\'e}s, La Paz, Bolivia.\\
$^{3}$Departamento de F{\'i}sica, FCFM, Universidad de Chile, Blanco Encalada 2008, Santiago, Chile.\\
$^{4}$Universidad de los Andes, Facultad de Ingenier{\'\i}a y Ciencias Aplicadas, Monse{\~n}or Alvaro del Portillo  12.455, Las Condes, Santiago, Chile.\\
$^{5}$Facult\'e des Sciences, Universit\'{e} Libre de Bruxelles (U.L.B.), CP\ 231, Campus Plaine, B-1050 Bruxelles, Belgium.\\}

\begin{abstract}
We propose here to interpret and model peculiar plant morphologies (cushions, tussocks) observed in the Andean altiplano as localized structures. Such structures resulting in a patchy, aperiodic aspect of the vegetation cover are hypothesized to self-organize thanks to the interplay between facilitation and competition processes occurring at the scale of basic plant components biologically referred to as 'ramets'. (Ramets are often of clonal origin.) To verify this interpretation, we applied a simple, fairly generic model (one integro-differential equation) emphasizing via Gaussian kernels non-local facilitative and competitive feedbacks of the vegetation biomass density on its own dynamics. We show that under realistic assumptions and parameter values relating to ramet scale, the model can reproduce some macroscopic features of the observed systems of patches and predict values for the inter-patch distance that match the distances encountered in the reference area (Sajama National Park in Bolivia). Prediction of the model can be confronted in the future to data on vegetation patterns along environmental gradients as to anticipate the possible effect of global change on those vegetation systems experiencing constraining environmental conditions.
\end{abstract}


\maketitle

\section{Introduction}

It is now widely acknowledged that interactions between plants can result in self-organized vegetation patterns, the scale of which is one to several orders of magnitude above the range at which an individual plant may exert facilitative as well as competitive influences on its neighbours ([1],[2]). The ranges of such influences are typically of tens of centimetres for herbs to meters for shrubs and trees, while the patterns observed at landscape scale may be characterized by a dominant scale over tens to hundreds of meters (Deblauwe et al. 2011; 2012 [3], [4]). Yet models of self-organization have shown that plant-plant interactions in resource (water, nutrients) deprived environments can indeed result in such landscape-scale patterns. The patterns that have so far received most attention display striking spatial alternation of vegetation and bare soil and a dominant wavelength can be detected ([5],[6],[7],[8],[3],[4]). Observed morphologies are made of bands, spots, labyrinths or gaps. Such patterns show a worldwide distribution ([9]) and are typically observed under arid or semiarid climates at the fringes of deserts, which suggest an analogy with phase transition between continuous vegetation cover (e.g. savanna) and desert. Broad scale monitoring of such systems thanks to remote-sensing has moreover shown that the periodic patterns do react to the variation of rainfall over decades in a way which qualitatively accords with predictions of the diverse self-organization models which have been proposed ([3],[4]). This provides ample evidence that those systems are not dependent on inherited geological or soil properties but are instead dynamical and react to ongoing rainfall variation, a fact that corroborate the hypothesis of self-organization under resource limitation.

However, biphasic mosaics associating vegetation and bare ground are a very widespread feature of vegetation cover in drylands ([10],[11],[12]), which is far more ubiquitous than the specific subclass of spatially periodic patterns. The processes from which such mosaics originate and perpetuate are qualitatively well understood and described by many authors: vegetation trap nutrients and help water infiltrate into the topsoil thereby creating 'fertile islands' within a resource deprived area ([13],[11],[12]). Conversely, the crusted surface of bare areas prevent water to infiltrate while nutrients, soil particles and plant propagules are moved along with water or wind towards vegetated places, which therefore concentrate the biological activity. In addition, the overall lack of resources leads plants growing in fertile patches extend their root systems under adjacent bare grounds thereby decreasing the already insufficient soil resources and reinforcing the contrast within the biphasic mosaic. This interplay between the positive feedback associated to vegetation and biological activity (i.e. facilitation) and the negative feedback represented by the competition for limiting resource (and/or the transfer of those resources towards vegetated places) qualitatively explain the emergence and perpetuation of biphasic mosaics thanks to self-organization processes. Several models have been designed to explain spatially periodic vegetation patterns (e.g.[14],[15],[16],[17]; see [18] for a review,[19],[20]). They have been inspired by the study of self-organization in physical and chemical systems ([21],[22]). They all conceptually feature the interplay between positive and negative feedbacks ([23]), while a necessary condition for stable patterns to occur is that positive processes are shorter-ranged (or implying lower diffusion) than negative processes ([24],[18]). 

But biphasic mosaics quite often correspond to aperiodic systems or even isolated structures. Trying to interpret such patterns as self-organized structures is nevertheless sensible since several models able to simulate self-organized periodic systems also yield aperiodic or isolated vegetation structures for particular domains in the parameters space. It therefore makes sense to investigate the origin and dynamics of dryland contrasted mosaics in reference to the physical concept of 'localized structures' ([25]). Indeed, recent applications proved fruitful ([26],[27],[28],[29]), although the concept so far barely percolated into vegetation science. For instance, circular bare ground spots in a continuous grassy cover (as observed in Namibia and therein called 'fairy circles') have been modelled as localized structures, and the model was able to yield sets of interacting circles as well as isolated ones (i.e. solitons) as is indeed observed in the field ([29]). Phase transitions at the interface between tropical grasslands and forests have also been addressed via the localized structure concept as to explain systems of isolated circular tree groves within seasonnally flooded, yet fire-prone grasslands ([26]). 

\begin{figure}[bbp]
\begin{center}
\includegraphics[width=15 cm,height=10.5 cm]{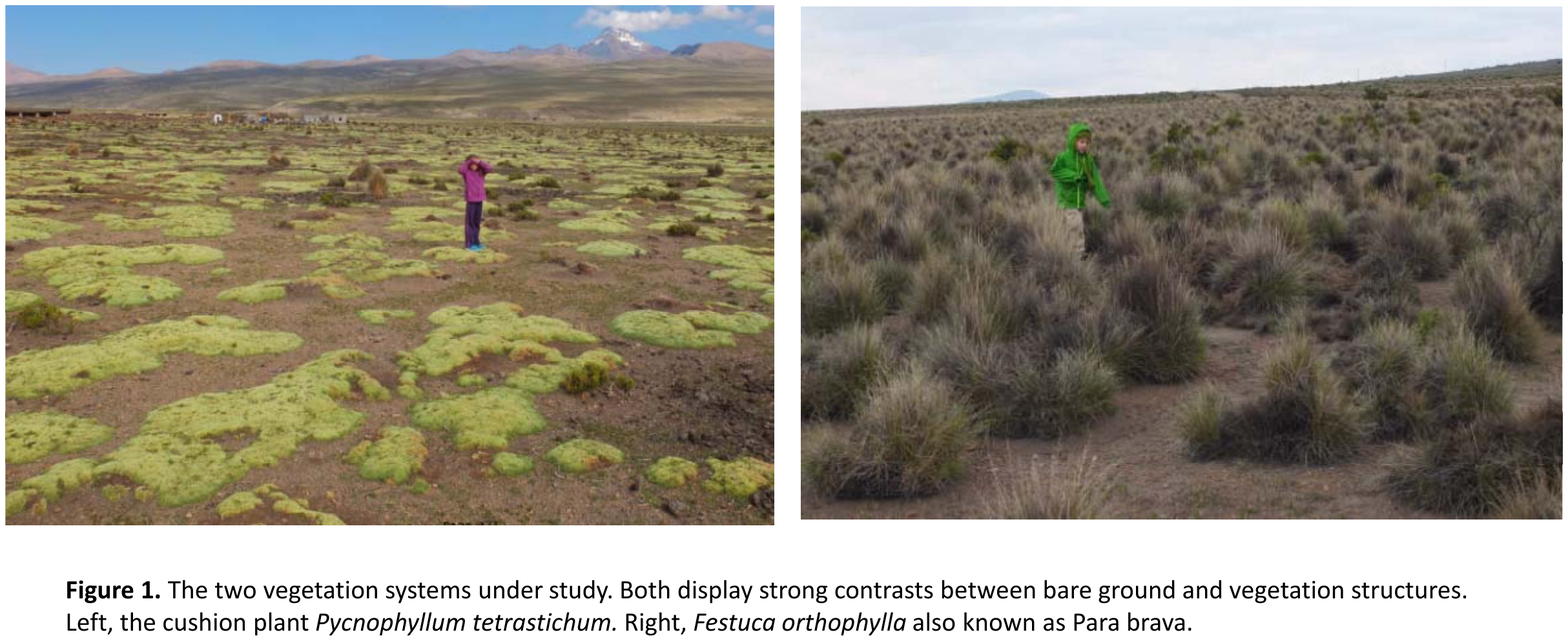}
\end{center}
\caption{The two vegetation systems under study. Both display strong contrasts between bare ground and vegetation structures. Left, the cushion plant \textit{Pycnophyllum tetrastichum}. Right, \textit{Festuca orthophylla} also known as Para brava;  Photos: F. Anthelme}
\label{Fig. 1.}
\end{figure} 

Besides systems which are spectacular because of the structure size that may reach 10-20 m ([26],[29]), there is however another very broad and ubiquitous class of vegetation patterns for which an interpretation as localized structures appeared relevant [27]. While occurring most frequently at small spatial scales (from tens of centimetres to several meters), it relates to a very pervasive feature of plants development, namely clonal reproduction. Many plants or plant structures are made of elementary components often referred to as 'ramets' that clonally duplicate and grow as to form larger vegetal entities, i.e. patches, which are frequently called 'genets'. Ramets are basic plant 'architectural' units, built upon one main axis (stem) and featuring leaves, roots and buds. The main ramet axis may ramify or extend (organogenesis vs. extension) as in any plant development scheme, but the most peculiar property of clonal plants, which is part of the broader category of reiteration processes ([30]) is that ramets replicate themselves many times as to extent the size of the patch (genet).  In doing so, genets may not only extend in area but also modify their overall shape. In two-dimensional space, this may lead either to the fractioning of the genet (for instance due to dieback of the central part) or to the possible merging of different genets ([31]. The first possibility leads to vegetation patches apparently distinct yet genetically homogeneous while the second outcome may result in vegetation patches of heterogeneous genetic identity [31]. Both cases contribute to an important consequence of clonal reproduction which is the blurring of the notion of 'individual plant'.

Clonal reproduction is particularly notable regarding plant forms of small size as grasses, herbs and shrubs. It also appears of increasing importance in ecosystems that are resource deprived and/or that experience strong climatic constraints (e.g. clonal propagation of Olea europaea subsp. laperrinei in the current climatic conditions in the Saharan mountains, [32]) or frequent disturbance (e.g. grazing,[33]). A pioneer work aiming at investigating clonal plant as self-organized localized structures was carried out in reference to a desert grass in the Negev desert ([27],[28]). In the present paper, we shall consider another constraining environmental context that corresponds to high altitude tropical drylands (tropical alpine regions) with a special reference to the Andes. Plants are there constrained by both cold temperature and a low rainfall-to-evapotranspiration ratio, while often experiencing grazing pressure from camelids (Llamas, Alpagas;[34]). In this environment, tussock forming grasses and cushions forming plants are important constituents of the vegetation cover [35]. In both cases (Fig. 1.), vegetation patches appear made of many tiny ramets that are called tillers in the case of grasses and rosettes for cushions ([36]). For each type of plant, we focus on a reference species, \textit{Festuca orthophylla} (Poaceae) for grasses and \textit{Pycnophyllum tetrastichum} (Caryophyllaceae) for cushions. Each species can locally dominate the vegetation within the study area by forming largely monodominant patches contrasting with bare soils (Fig. 1.; see also [37],[38]). 

We hypothesize that such patches can be modelled as self-organized localized structures that are dynamical outcomes of the interactions occurring between ramets. Competitive interactions stem from the need for each ramet to track insufficient soil resource by developing lateral root systems (Fig. 2.)thereby overlapping with the area that neighbouring ramets need to prospect. Facilitative interactions result from the influence that vegetation exerts on the physical environment by mitigating harsh temperature, reducing evaporation and improving water infiltration (e.g. [39]). Such facilitative influences are extensively described in the literature [40], and more specifically for the two particular plant systems we are dealing with (see [41] and [38], regarding \textit{F. orthophylla}, and [42],[43], regarding cushions).

We aim here to model and interpret clonal structures in the Andes as localized structures by referring to the modelling framework introduced by Lefever and Lejeune ([14]) and further developed by Lefever et al.[44] (see also [1]). This framework models the dynamics of the plant biomass using a single integro-differential equation that features logistic growth and non-local modulations expressing facilitative and competitive feedbacks of the biomass on its own developments. Such modulations are modelled via kernels that directly relate to plant morphology: the kernels embodying competitive and facilitative modulations have ranges of influences commensurate with the dimensions of the plant rooting system (rhizosphere) and above-ground part, respectively. In order to consider the dynamics of clonal systems, the emphasis will be here shifted from the plant to the ramet and the ranges of facilitative and competitive influences assessed accordingly. 

\begin{figure}[bbp]
\begin{center}
\includegraphics[width=14 cm,height=8 cm]{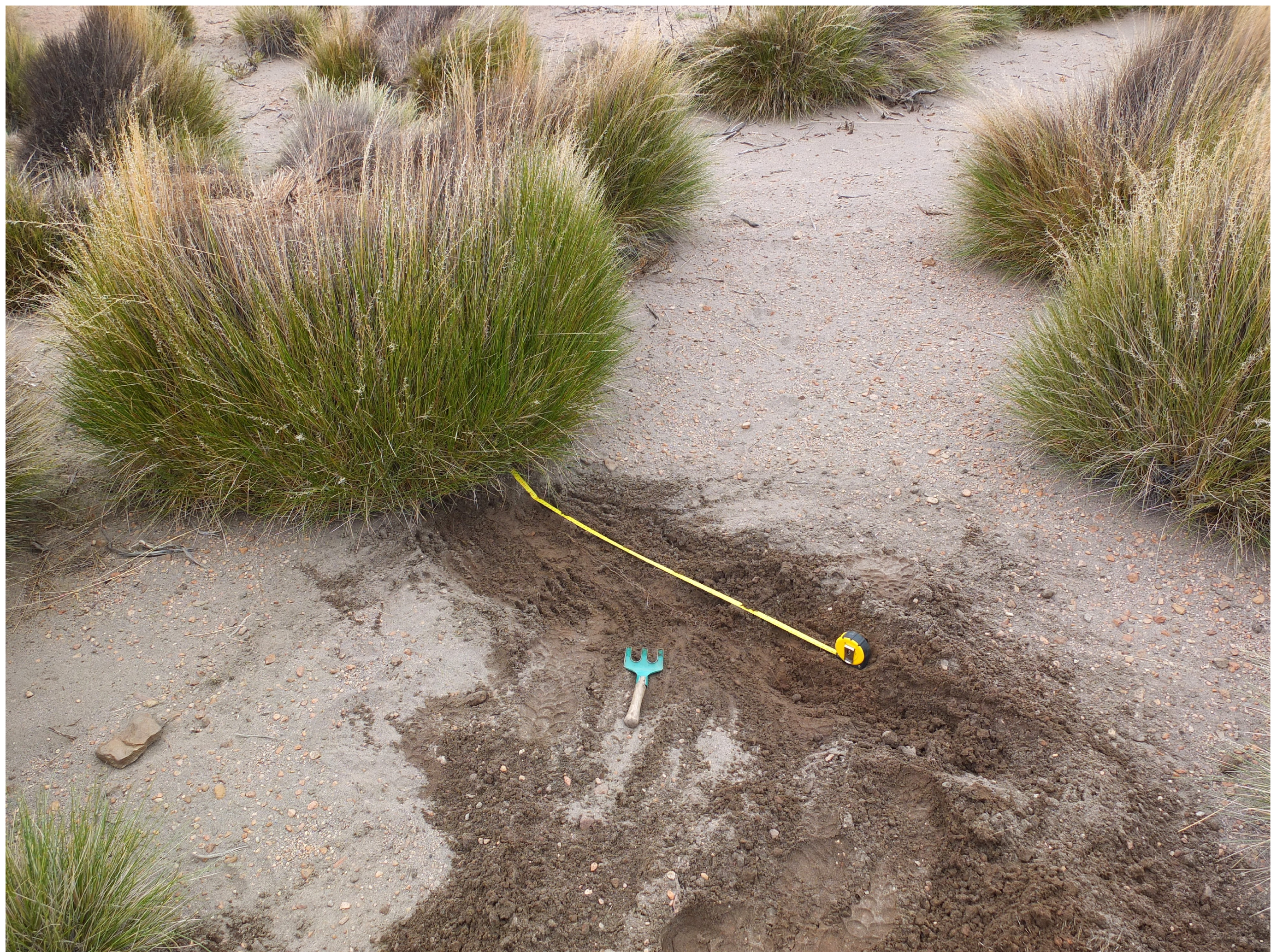}
\end{center}
\caption{Lateral extent of the superficial rooting system at the periphery of the two plant forms under study.  A small tussock of \textit{Festuca orthophylla} \textit{Pycnophyllum tetrastichum.  Photos: F. Anthelme}. The measuring tape parallels one of the lateral root and indicate 35 cm .  Note that the boundaries of the tussock/cushion have been preserved during root excavation.}
\label{Fig. 2.}
\end{figure} 

In this paper, we will specifically refer to the version of the model proposed by [44] in relation to the modelling of spatially periodic bare grounds areas or 'gaps' punctuating a vegetation matrix of comparatively high biomass (i.e. 'deep gaps'). This model was also applied in [29] to the modelling of circular gaps or 'fairy circles' [45] occurring as isolated structures as well as aperiodic systems of gaps within a continuous grass matrix as observed in the drylands of Namibia. It demonstrated the potential of the model to produce localized structures and to predict some of their characteristics. Here we will explore the potential of the model to account for localized structures occurring as 'bumps' instead of 'trough', as to render clonal vegetation patches isolated within a bare soil matrix. We will show that considering the available data on ramet morphology and biomass dynamics in the reference ecosystem, the model is able to predict macroscopic features of the observed patterns (e.g. tussock size and modal inter-tussock distance). We will carry out a thorough interpretation of the available information as to provide realistic values for the model parameters. On this basis, we intend to show that using this array of parameters values relating to processes at ramet scale, the model is able to provide predictions and at the scale for which plant biomass display the maximum level of spatial variation (as illustrated in Fig.1.)

\section{Mean-field model for vegetation dynamics}
It is well known in literature that the concept of self-organization based on the idea of Turing ([21]) then Prigogine and Lefever ([22]) involves large spatial scales, in the sense that the macroscopic spatial scale resulting from a self-organization could be several orders of magnitude larger than the size of individual elements such as atoms, molecules, or trees, which interactions can be at the basis of self-organization. This universal process has been experimentally proved in several physical and chemical systems including optics and lasers (see recent overviews on this issue: [46],[47],[48],[49]). The self-organisation mechanism was applied to the context of plant ecology as to explain spatially periodic vegetation patterns displaying large scale periodicity compared to individual plant sizes and the ranges of spatial scales at which the elementary processes resulting from plant-plant interactions may take place in the field [1]. For instance patterns involving grasses may reach a dominant wavelength of up to10 m. When shrubs and trees are involved, the order of magnitude of the wavelength is often in the range of 40 m to 100 m and may exceed 120 meters [4]. 

Several families of models have been proposed to deal with self-organization in vegetation (see [18]for a review). All conceptually relate to the interplay between facilitative and competitive feedbacks of vegetation biomass on its own dynamics acting at distinct spatial scales [23]. We here refer to the modelling framework introduced in the seminal paper by Lefever and Lejeune [14] and more precisely to the development proposed in [44]. This variant of the model has been specifically applied to the modelling of 'deeply gapped vegetation' often observed in the sub-Saharan Sahel where patches of bare grounds regularly punctuate shrubby vegetation of comparatively high biomass. In a first application, [50] assessed the parameters of the model in reference to deeply gapped vegetation observed in Niger and discussed in depth the meaning of the parameters while confronting field measurements to theoretical predictions. 

The model consists of a single logistic equation governing the dynamics of the vegetation biomass density through the balance between biomass growth and death. The growth and  death processes are modeled by the following logistic equation governing the evolution of the vegetation density $b=b({\bf {r}},t)$ at time $t$ and at the point ${\bf {r}}=(x,y)$. 
\begin{equation}
\partial_t b=b\left(1-b\right)\mathcal{M}_f-\mu b\mathcal{M}_c +D \mathcal{M}_d \end{equation}
The first term  expresses the rate at which the biomass density increases and saturates. This is  the biomass 
gain that  corresponds to the natural production of plants via seed production, dissemination, 
germination and development of shoots into new mature plants.  The second term models the 
biomass losses which describes death or destruction by grazing, fire, termites, or herbivores. 
The parameter $\mu$ measures the resources scarcity, and more generally the environmental adversity. When the water becomes scarce, plants adapt their roots systems to fight against water scarcity. They strive to maintain their water  uptake by increasing the length of their roots (Fig. 2.). They thus compete with other plants on long distance $L_c$. This is a negative feedback that tends to reduce the biomass density modeled by the function $\mathcal{M}_c$. 

\begin{figure}[bbp]
\begin{center}
\includegraphics[width=10cm,height=10cm]{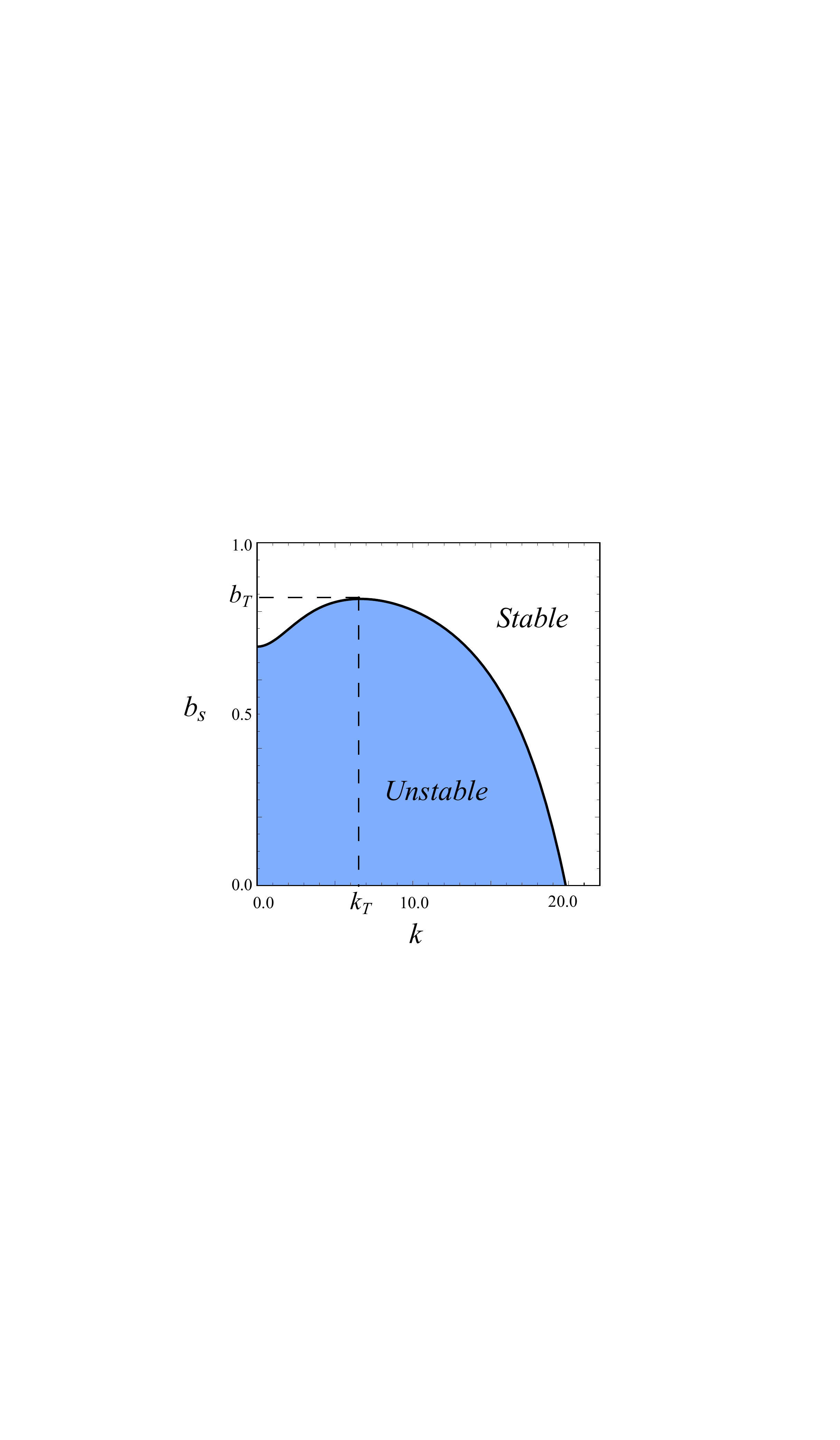}
\end{center}
\caption{Marginal stability curve for the homogeneous steady state solutions. Parameters used are: $\xi_{c}=2 $, $\xi_{f}=5 $, $L_c=0.4$, $L_f=0.1$ the domain of instability is represented by a light blue area in the $(b_s,k)$-plane. $b_s$ is the homogeneous steady state biomass density. The stability vs.instability domains are separated by a solid line. $L_T$ and $k_T$ correspond to the threshold at which the symmetry breaking instability appears.}
\label{Fig. 3.}
\end{figure} 

In the following we focus on the Gaussian type Kernel for both competition and facilitation, i.e.,
\begin{equation}
\mathcal{M}_{c,f}=\exp\left[\xi_{c,f} \int K_{c,f}(|{\bf r^{\prime}}-{\bf {r}}|) b({\bf r^\prime},t)d\bf r^{{\prime}}\right],   \text{with } K_{c,f}(|{\bf r^{\prime}}-{\bf {r}}|)=N_{c,f}\exp\left(-\frac{|{\bf r^{\prime}}-{\bf {r}}|^2}{L_{c,f}^2}\right)
\end{equation}
where $N_c$ is the normalization coefficients are
\begin{equation}
N_{c,f}=\frac{1}{\int \exp\left(-\frac{|{\bf {r}}|^2}{L_{c,f}^2}\right)d{\bf r}}
\end{equation}
facilitative interaction between plants is modeled by the  function $\mathcal{M}_f$ which expresses the positive feedback which favors the vegetation development.  The parameters ${\xi_{c}}$ and ${\xi_{f}}$ model the interaction strength associated with the competitive and facilitative processes, respectively.  The last term expresses the vegetation spatial  propagation through seed dispersion, production and germination.
 \begin{equation}
\mathcal{M}_{d}=\int K_{d}(|{\bf r^{\prime}}-{\bf {r}}|) \left(b({\bf r^\prime},t)-b({\bf r},t)\right)d{\bf r^{{\prime}}},   \text{with } K_{d}(|{\bf r^{\prime}}-{\bf {r}}|)=\frac{\sigma}{\pi}\exp\left(-\sigma|{\bf r^{\prime}}-{\bf {r}}|^2\right)
\end{equation} 
 Finally the homogeneous steady states of Eq. (1) are the branch of trivial solutions $b_=0$ that represents territories devoid of vegetation: $b_=0$ is unstable for $0<\mu<1$  and stable for $\mu >1$. The two other homogenous solutions are
 \begin{equation}
\mu=(1-b_s)\exp (\Lambda b_s).
\label{HSS}
\end{equation} 
 They represent spatially uniform plant distributions. For $\Lambda>1$ and $1\leq \mu \leq\exp (\Lambda-1)/\Lambda$, Eq. (\ref{HSS}) admits two nontrivial solutions $b_{s-}$ and $b_{s+}$. The uniform state $b_{s-}$ is always unstable. 
 
 We now look for the conditions under which spatially uniform distributions of vegetation  $b_{s+}$ is  unstable with regard to inhomogeneous perturbations. In the Fourier representation, growing modes are characterized by a finite interval of wavenumbers. This Turing kind of symmetry-breaking instability  produces patterns characterized by an intrinsic wavelength determined by the system's dynamics rather than by geometrical factors and/or boundary conditions.
Small amplitude deviations from $b_{s+}$ in terms of Fourier modes  in the space of wavevectors $\exp (i{\bf k} \cdot {\bf r}+\lambda t)$ are considered. This analysis yields the eigenvalues of the linear operator
 \begin{eqnarray}
\lambda&=&(1-2b_s)\exp({\xi_{f}b_s})+(b_s-1)b_s\xi_{f}\exp\left(\xi_{f}b_s-\frac{k^2L_f^2}{4}\right)\nonumber \\
&-&\mu \exp(\xi_{c}b_s)\left (-1+b_s\xi_{f}\exp (-\frac{k^2L_c^2}{4})\right )+D\left (\exp (-\frac{k^2\sigma}{4})-1\right )\label{Spectrum}
\end{eqnarray}
The wavelength of the first non-zero-Fourier mode to become unstable is 
\begin{equation}
\lambda_T=\pi\sqrt{\frac{L_f^2-L_c^2}{\log\left(\frac{L_f^2\xi_{f}}{L_c^2\xi_{c}}\right)}}.
\label{wavelength}
\end{equation} 
The threshold state at which the symmetry breaking instability appears on the $b_{s+}$ branch  of solutions is
 \begin{eqnarray}
 \left (1+b_T\xi_{c}\exp (-\frac{k_T^2L_c^2}{4})\right )&=&b_T(b_T-1)\xi_{f}\exp\left(-\frac{k_T^2L_f^2}{4}\right)+(1-2b_T)\exp({\xi_{f}b_T})\nonumber \\ 
 \text{with   } \mu_T&=&(1-b_T)\exp (\Lambda b_T)  \text{   and    } k_T=\frac{2\pi}{\lambda_T} 
\label{threshold}
\end{eqnarray}

The domain of instability includes the wavenumber of the fastest-growing modulation . The critical wavenumber
corresponds to the situation where the homogeneous steady state $b_{s+}$ exhibit a pattern forming (Turing) instability as shown in Fig. 3. The coordinates in parameter space at which the symmetry-breaking instability takes place are $b_T=0.78$ and $k_T=4.6$. The corresponding aridity parameter at this instability is $\mu_T=2.28$ and the wavelength $\lambda_T=2\pi/k_T=1.36$ m.

\begin{figure}[bbp]
\begin{center}
\includegraphics[width=13cm,height=8.cm]{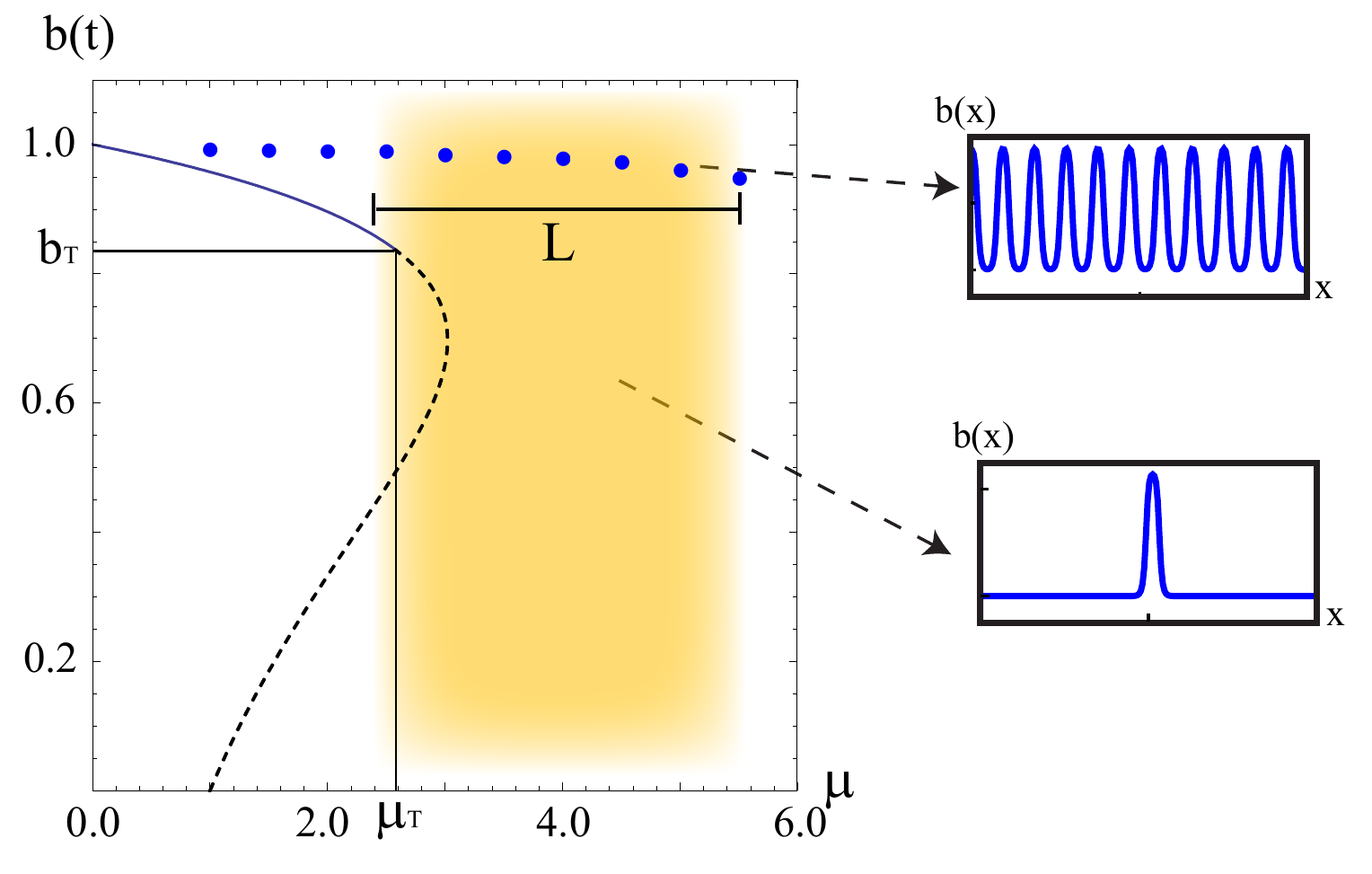}
\end{center}
\caption{Bifurcation diagram. Same parameters as in Fig. 3. The value of the parameter $\mu$ which is indicated and used for simulations is 2.7. The solid and dashed lines correspond to stable and unstable homogenous steady states, respectively. The blue dots are the maximal biomass density values obtained at the centre of the localized structures for the one-dimensional simulations of the patterns (illustrations on the right side of the figure) with increasing $\mu$ values. The domain where localized solutions exist is denoted by L.}
\label{Fig. 4.}
\end{figure} 

\section{Reference vegetation patterns and parameter assessment:}

Large areas experiencing tropical alpine environments (TAE) are encountered in Africa and South/Central America while 90\% of these environments is observed in the Andes [51]. TAE experience cold minimum temperatures that constrain plant development [52], along with some distinctive features: (1) an inversion of rainfall gradients in the form of increasing aridity at higher elevation, (2) strong nyctemeral variations (exceeding annual variations) in temperatures (3) the absence of persisting snow cover contrary to most alpine environments ([52],[53]). In the present study, we will specifically refer to the Sajama National Park (SNP) in Bolivia for which climatic features, as exposed in [54] fits to the overall descriptions of TAE. Notably, the annual precipitation is around 350 mm and is concentrated between November to March. Locations in the SNP where direct observations have been carried out by one of us (F. Anthelme) experience the same climate.

The singular characteristics of TAE make some plant life forms specific to these regions, or at least let them be found there at higher abundance. Among them, cushion-forming plants (hereafter termed cushions) are incredibly diverse and may reach giant sizes. Their abundance can reach 50\% of the soil surface in dry alpine areas where vegetation is fragmented (see Fig.1.), notably in the Chilean high Andes [55]. However, the relative cover of cushions in dry areas never approaches 100\% and the patterns are biphasic. The 3D shape of cushions is variable, but in general it can be described as relatively flat, with a low growth rate. The facilitative effects of cushions on plant communities in dry alpine environments are well known [55]. Cushions ameliorate minimum temperature, water and nutrient availability. Though intraspecific facilitation within cushion populations and between ramets of cushions has not been tested explicitly so far, it is highly probable that their structure, often composed of a large number of small rosettes (ramets) generating a compact cover [36], determines intense interactions between neighbouring rosettes, both positive and negative. Tussock grasses are frequently observed and often a dominant life form in TAE ([36],[56],[57]). There, tussock grasses display a typical two-phase structure with tussocks having a limited lateral growth and a high density of tall stems at tussock centre (Fig. 1.). Although providing more humidity and nutrient than available in adjacent bare areas (see below), tussocks in dry environments are also thought to have a negative impact on other plants, because of strong competitive traits [58]. 

\textit{Interpretation of the tussock pattern in the SNP:}
We here equate $b$ (Eq.2.1.) with the above-ground biomass density (referred to hereafter as 'biomass') irrespective of the below-ground part. The density is defined with respect the maximal local biomass observed, which is generally observable in the middle of a vegetation patch. In the case of \textit{F. orthophylla}we assume accordingly that a density of 1 is reached in the centre of every mature, non-senescent tussock, while the density is decreasing towards the periphery as described in [41].  For simplicity, the model does not distinguish the 'live' and 'dead' fractions of the biomass as in [41] and $b$ can also be considered as a 'phytomass' sensu [41]. The average biomass density over a tussock is estimated of ca. 0.5. The overall basal area of the tussocks is reported to be of ca. 15-20\%, a range of values which appears stable across space from both literature [41] and direct measurements from one of us carried out in the SNP. We therefore assess the overall biomass density of this typical pattern as $\tilde{b}=0.18*0.5$, i.e. ca. 0.1.

In the model, non local modulations express how the balance between biomass build up and decay at the scale of a given ramet is affected by the influences, either competitive or facilitative exerted by neighbouring ramets. Associated parameters are, first, the ranges of interactions ($L_f$ and $L_c$, i.e. the ranges of the Gaussian kernels) which reflect ramets morphology and, second, the intensity of the modulations (expressed by $\xi_f$ and $\xi_c$) of the overall biomass dynamics as ruled by $\mu$.

\textit{Morphological parameters:}
Based on the available published data and direct field observations (Fig. 2.) we set $L_c$ to the observed length of lateral roots (i.e. 40 cm for both the observed plant forms). We assessed the range of the facilitation effect resulting from lateral shading (integrating the effect of dusk and dawn sun inclinations) and protection from grazing as approximately half the height of an average mature ramet. In the case of the \textit{F. orthophylla}grass, the average height of live tillers leaves is of about 20-30 cm, and in accord with the modelling of interactions through Gaussian kernels, we set $L_f$ to 10-15 cm. 

\textit{Facilitative and competitive feedbacks from the aboveground biomass}
The intensity of facilitative and competitive modulations (parameters $\xi_f$ and $\xi_c$) is first assessed in reference to the influence of vegetation on soil moisture (considered via the volumetric soil moisture content, vol\%) within the shallow rooting zone. In the topsoil that is actually explored by tussock roots (< 40-50 cm), soil moisture proved to be virtually undetectable during the dry season whatever the location [41]. But during the rainy season, strong variations were observed in time (depending on the occurrence of rain showers) as well as in space (under vegetation cover vs. in open ground). Maximal values of soil volumetric moisture were reported to be above 5.5\% under a grass tussock while they never exceeded 1.5\% in the absence of vegetation. Moreover, a mulching experiment let soil moisture content reach as much as 19.4 [41], emphasizing the pervasive role of evaporation in depleting the scarce soil water resource. Moreover, the latter authors reported that the tussock aerial part was at least as efficient as the mulch to decrease the hottest mid-day soil surface temperature and henceforth the evaporative demand. Hence, reinterpreting the published results allowed us to separate the relative effects of soil water consumption by grass tillers (competition) from the protection against evaporation (facilitation). It thereby appears that the reduction of evaporation by grass cover represents a possible increase in maximal volumetric soil moisture content up to 12 times the maximal content found in bare areas. Conversely, the potential water demand from well developed tillers represents ca. 9 times the bare soil content. The overall balance appears thus positive thereby illustrating the predominance of facilitation over competition that helps tillers survive adverse climatic conditions as soon as they are part of a tussock. 

From those figures it looks as if at the centre of a modal tussock the soil moisture was ca. 2.5 times the level it reached under bare ground, far away from any vegetation influence. Reasoning on moisture values averaged over a rainy season instead of maximal values yields a lower, yet positive figure of ca. 1.5 under vegetation. It is however reasonable to think that maximal values better render the functioning of the system which is probably based on pulses of vegetation development triggered by discrete rainfall events. For instance, [50] described the functioning of a semiarid system where the most striking difference in terms of water budget between vegetated and bare locations was mostly noted during the days following a rain shower before progressively fading away as the topsoil water dries up everywhere. 

To traduce the above values into the parameters of the model (i.e., $\xi_f$ and $\xi_f$), we can refer to the effect of unusually low rainfall, as reported by [41] who compared biomass production between 2007-2008 (considered as a 'normal' rainy season, 384 mm) and 2006-2007 (poor season, 327 mm, minus 15\%). As a response of \textit{F. orthophylla} to the decrease in rainfall, the mean standing tussok biomass dwindled by 40\% (Table 5 in [41]). But so strong a decrease was not a direct result of tillers death but rather reflected a reduction of the green fraction in the well developed tillers[41], and caution is needed when interpreting it. On the other hand, publications from water limited ecosystems with annual rainfall around 350 mm per year reported that grass biomass reduction accompanying a 15\% rainfall decrease is of about 30\% [59]. From this observation and assuming a linear response of biomass production to any water resource fluctuations of reasonable amplitude, we set a ratio of 2 (0.3/0.15) and apply it to the modulation in soil moisture determined by a tiller bunch of modal size, as described above. According to the model, this effect is to be equated to $exp(\xi_f b^t)$, for moisture increase and $exp(\xi_cb^t)$, for moisture decrease, where $b^t$ is the mean biomass density of tillers over a tussock that is ca. 0.5. This yields for the two parameters, $\xi_f= 6.3$ and $\xi_c =5.8$.
$\Lambda$ therefore amounts to ca. 0.5 making the vegetation system look as weakly cooperative considering the only aspect of vegetation feedbacks on biomass production through soil water resource. This weakness of the positive balance is due to the strong demand of the grassy cover for transpiration. However, there is another facilitative feedback of plant biomass that acts as a protection against grazing. It can be considered as additive since it is independent from the soil water resource. We will assess it in the subsequent point as part of the overall biomass dynamics.

\textit{Overall biomass dynamics:}

$\mu$  is basically the ratio of the plant biomass lost to the biomass produced at low biomass levels (i.e. far under the logistic saturation). In the considered ecosystem, biomass loss features two components, which are senescence of live plant material as well as live biomass destruction by grazing. Those two causes each determine additive fractions of $\mu$, i.e.  $\mu =  \mu_0 + \mu_g$  relating to senescence and grazing, respectively.

$\mu$ fundamentally expresses the potential development of a small amount of live plant biomass (typically a small bunch of young ramets) that does not benefit from facilitative retroaction from pre-existing developed plants. The young plant will either die out if the conditions are too unfavourable ($\mu < 1$) or develop at a rate which is a decreasing function of $\mu$. Qualitatively, we expect $\mu$  to be a decreasing function of rainfall. From experimental results [41] and direct field observations, it appears that without any facilitative effects by large preexisting biomass tiller development is very weak and may not completely balance decay. So we set $\mu$ to ca. 1.5.

In the case of \textit{F. orthophylla} we shall rely on an experiment reported by [60] (p.31) that provides elements to assess the grazing-related part of $\mu$: after experimental burning (i.e. removal of most of the above-ground biomass), the biomass regrowth is dependent on the root system which has been spared by fire. In fenced plots that were not accessible to camelids, the total biomass production during the first year of regrowth proved seven times higher than in unfenced plots subjected to grazing. Assuming an exponential regrowth curve at plot scale, since the system restarts from virtually no above ground biomass, it yields: $\mu_g=log(b_f/b_u)/(t_f)$, where $b_f/b_u$  is the ratio at the end of the first year after burning of biomass in fenced to unfenced plots and $t_f$ is one year or ca. 2 tiller generations [41]. Accordingly, we assess $\mu= log(7)/2 =1$.

Another aspect of the same experimental design also showed that fencing out camelids only determined a small increase in biomass production (ca. 10\%) at and around the biomass levels of the patterns usually observed in the studied area, i.e. $\tilde{b}$. This indicates that shielding a large share of young and productive tillers against grazing is an important component of the positive feedback that a grass stand exerts on its own dynamics. It is apparently as if this modality of facilitation cancels out a very large share of the grazing impact. To assess the magnitude of such a protecting effect and deduce the final value of $\xi_f$, we relate the slight increase in biomass production $\Delta_s$ between fenced and unfenced situations to $\mu_g$ and $\xi_f^0$ as :
\begin{equation}  
\xi_f-\xi_f^0=\frac{exp{(-\xi_f^0b)}}{b(1-b)}\left(\mu_g+\Delta_s/b\right)
\end{equation} 

Where $\xi_f^0$ and $\Lambda^0$ (6.3 and 0.5, respectively) correspond to values assessed at the previous point and refer to the only aspect of vegetation feedback on soil moisture.  This finally yields  $\xi_f=9.1$ and $\Lambda=3.3$. We will therefore use an order of $\Lambda=3$  for further computations and simulations.

\section{Clustering and localized vegetation patterns}
A bifurcation diagram in the plane ($b$-$\mu$) is shown in Fig. 4. The solid line corresponds to stable homogeneous steady state (HSS) solutions, doted lines corresponds to unstable HSS. The circles indicate the maximum values corresponding to the periodic structures that emerge sub critically from the Turing instability located at $b_T$, $\mu_T$. We focus on the regime where the system exhibit a coexistence between the bare stable $b_0$ and the periodic vegetation pattern. In this region of parameter there exists a domain of parameters denoted by $L$ where stable localized vegetation patches are stable. Examples of stable localized vegetation patches with one, two or several peaks are shown in Fig. 5. The numbers of peaks in the localized pattern are determined solely by the initial conditions used. All this localized structures are obtained for the same parameters values. They can either be self-organized or randomly distributed in space as shown in Fig.5. Localized structures or localized patterns are homoclinic solutions (solitary or stationary pulses) of partial differential equations such as reaction-diffusion models or integro-differential equations such as Model Eq. (1). The condition under which localized structures and periodic patterns appear are closely related. Typically, when the Turing instability becomes sub-critical, there exists a pinning domain where localized structures are stable. They occur in various fields of nonlinear science, such as chemistry ([61],[62],[63],[64],[65]) and optics ([25],[66],[67]). 

For the tussock grass, we use the parameters estimated above to compute the wavelength expected for the pattern at the first non-zero Fourier mode to become unstable (equation 2.7).  It yields 0.95 m, i.e. within the range of distances between tussock centres measured in the field (0.8 to 1 m). $L_T$ strongly depends on the values taken for the ranges of facilitative and competitive interactions between ramets. Simulations that took $L_f = 0.1$ m and $L c = 0.4$ m, also reached a wavelength of ca. 1 m. This indicates a certain robustness of the result against variations of the most influential parameters within a realistic range of uncertainty.

\begin{figure}[bbp]
\begin{center}
\includegraphics[width=15cm,,height=8.5cm]{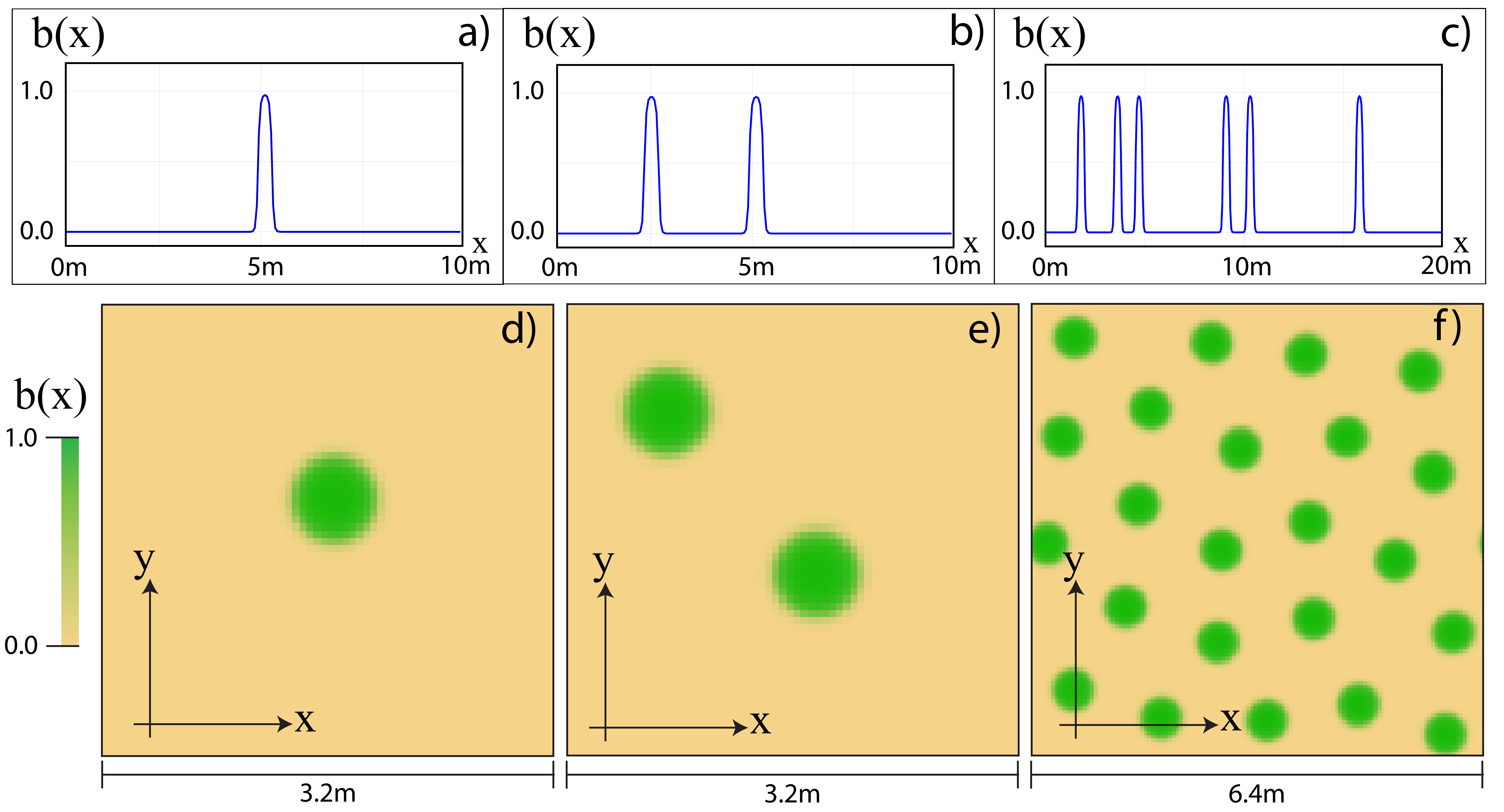}
\end{center}
\caption{Simulated structures. The figure shows the variation of the number of localized structures according to the initial conditions used for simulations, in one dimension: a), b) and c) and in two dimensions: d), e) and f), note that the two series are independent. All insets display the variation of biomass density in 1D or 2D, and the results have been obtained for the same set of parameters values as in Fig. 4, with $\mu=2.7$.}
\label{Fig. 5.}
\end{figure} 

\section{Discussion-Conclusion}

We have here proposed interpreting patchy patterns displayed by herbaceous, clonal vegetation in the tropical alpine regions of the Andes as self-organized localized structures (LS), in environments for which vegetated cover and bare soil are simultaneously stable. Several elements encouraged us towards this type of interpretation. First, it is widely acknowledged from botanical observations that elementary plant units or ramets play a fundamental role from both the dynamical and functional points of view in what appears in Fig.1. as 'vegetation patches' (e.g. [68],[69]). Moreover, there is for many species no information on what the genetic structure of the patches may be and there is no a priori reason to consider the overall dynamics of a given patch as constrained or influenced by an overall genetic program as for some other plant forms such as trees or annual herbs. For example, the subantarctic cushions formed by \textit{Azorella selago} were demonstrated to be the result of the aggregation of several individuals with distinctive genetic material, and not only one single genet ([31]). Several qualitative observations support such an interpretation. Patches are not permanent entities since they are known to change shape and even divide as part of their dynamics. Although we did not put emphasis on this aspect, senescence of the central part of the tussock grass \textit{F. orthophylla}is also observable in our reference area, as was described and modelled for other grass species in various types of ecosystems, e.g. \textit{Poa bulbosa} in the Negev desert [28] or \textit{Scirpus holoschoenus} in Eastern Central Italy [70]. Clearly, such dynamics casts doubt on considering patches as 'individuals'. Moreover, for certain species, the planar growth of a patch may appear as unlimited once the resource constraint and/or the competition from other species has been removed or alleviated. We may here quote the case of Andean cushions bogs, especially those dominated by \textit{Distichia muscoides} (Juncaceae), a cushion plant found in the tropical alpine environments of the Andes: on dry slopes it forms small, circular cushions while in wet bottomlands with saturating water conditions it can develop a continuous cover ([71]; Anthelme pers. obs.). It therefore appears sensible to look at those patches strongly contrasting with bare ground as structures emerging from the interactions between the ramets in the context of environmental conditions unfavourable to plant growth. 

A large share of self-organization models that have been developed to account for spatially periodic vegetation patterns can also yield LS within the part of the parameter space that corresponds to conditions harsher than for periodic patterns. Under such adverse conditions, plant LS owe their existence and perpetuation to strong facilitative effects from pre-existing biomass and the locations of the structures are strongly dependent on initial conditions. With the model [44] on which we relied here, the adversity of the environment is measured by a single parameter $\mu$, which should strongly exceed 1 (the level at which biomass production and destruction are balanced) if LS are to be obtained. In addition, facilitative effects should markedly exceed competitive influences and a strongly positive value of $\Lambda$ is expected. In the case, of \textit{F. orthophylla}, reinterpreting soil moisture data from [41] allowed us to assess $\Lambda$ via the modulation coefficients $\xi_f$ and $\xi_c$. It appeared however that the influence of the pre-established tussocks on the soil moisture, though globally improving the soil moisture budget does not yield strong enough values of $\Lambda$ to match very strong values of $\mu$. On the other hand, acknowledging $\mu$ as cumulating both aridity and grazing as additive sources of environmental adversity led to a strong overall value of $\Lambda$ (ca. 3). Grazing is indeed a crucial component of these ecosystems, as suggested by the experimental results of [60]: fencing out camelids appeared necessary to let the above-ground biomass rebuild from the root systems during the first year after burning. This experiment also qualitatively confirmed that the development of young ramets crucially needs the sheltering of extant biomass. The pervasive role of facilitation ([39]) is here illustrated all the more that growing from preserved root systems is far less demanding than growing from isolated seed germination, i.e. virtually from scratch with no facilitation. 

\begin{figure}[bbp]
\begin{center}
\includegraphics[width=18.5cm,,height=10.5cm]{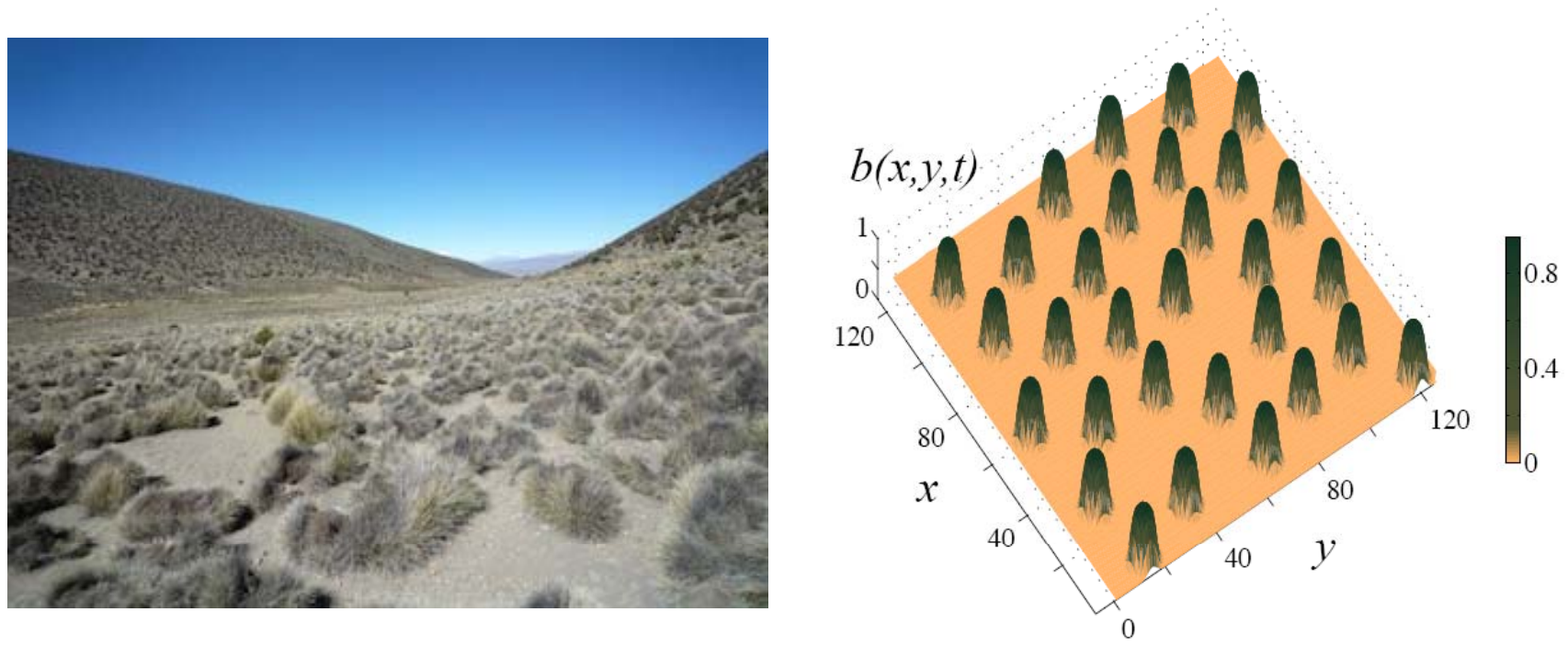}
\end{center}
\caption{Overall comparison of a system of tussocks of \textit{Festuca orthophylla} (observed in the Sajama National Park, Bolivia) with the simulation results. The simulated system displays a mean distance between tussock centres that agrees with the range of values observed in the field (i.e. 0.8 to 1 m).}
\label{Fig. 6.}
\end{figure} 

Hence facilitation appears here pervasive in the case of ramets belonging to a same species, while there is a growing body of literature emphasizing the role of inter-specific facilitation in tropical alpine ecosystems vegetation (see [53] for a review). On the other hand, a fencing out experiment when applied to unburned tussock patterns of current biomass density result only in a modest biomass increase in the ungrazed compartment [60]. This pleads for strong constraints endogenous to a mature system that reflect competition for limited soil resources and even for space: the central parts of the tussocks are at saturating biomass levels (as rendered by the logistic term in the model), while at the periphery, competition from extended lateral root systems is blatant (Fig. 2.) and hinders ramets development at the fringes. (Besides, it is highly probable that not only soil moisture but also nutrients are limiting; see p.31 in [60]) As a consequence, the removal of the grazing pressure does not trigger a rapid development of the overall biomass of the system, which can however be considered of low density $\tilde{b}$ with respect to the maximal biomass encountered at the centre of most patches (ca.1): $\tilde{b}$ was assessed in the range 0.08 to 0.15 since the basal area of the tussocks covers around one fifth to one sixth of the ground surface. 

Low biomass density, strong facilitative and competitive effects and overall sensitivity to the initial distribution of biomass, all those conclusions are consistent with the interpretation of the patchy vegetation patterns on the Altiplano as systems of LS. In this paper, we aimed to verify that interpretation by matching predictions of a simple, fairly generic model of vegetation dynamics with some macroscopic features of the observed systems. In the best documented case of \textit{F. orthophylla}(thanks to the considerable work carried out by JAF Monteiro [60]), both our analytical predictions and simulations yielded an inter-tussock distance $L_T$ within the range of values observed in the field (i.e. 0.8 to 1 m). Moreover, the equation predicting $L_T$ appears mainly sensitive to the ranges of facilitative and competitive interactions, i.e. $L_f$ and $L_c$. Considering the morphology of cushions (Fig. 2.), similar values of $L_T$ are probable for this life-form, and indeed field measurements in the location where the picture of Fig. 1. was taken pointed towards mean $L_T$  values of 1.4 m

In this paper, we show that the systems of vegetation patches made of clonal plants and observable in the tropical alpine region of the Andes are plausibly patterns self-organizing under the form of localized structures. Yet, to progress in our understanding of the phenomenon, we need additional observations and data at both landscape and plant scales. At broad scale, it would be useful to acquire aerial photographs of patterns (for this, drones now provide an efficient solution) as to check patterns characteristics over rainfall gradients. In the Sajama National Park, patterns look aperiodic while patches appear to have a modal size (Fig. 1.), but this is to be systematically checked in diverse locations and conditions. Image analyses of patterns over ecological gradients proved fruitful in other contexts ([6],[8],[3],[72]). If our interpretation based on LS is correct, LS systems may shift towards periodic structures under wetter climates. Conversely, in drier or more heavily grazed situations, systems of LS are expected to give way to more scattered LS of highly random spatial distribution (as for simulations in Fig. 5). \textit{F. orthophylla}is obviously a good opportunity for such analyses, especially because it is observed at regional scale of the Andean Altiplano along a gradient of increasing aridity,  from Lake Titicaca to the Salar of Uyuni (Anthelme, pers. obs.). At ramet scale, and since our reference model ([44]) emphasized plant morphology, more accurate values of interaction ranges should be deduced from an analysis of how ramets ramify and how bunches of ramets of common origin connect to the longest lateral roots (Fig. 2.) that determines the competition range. Greater accuracy in model predictions, via enhanced measurements of interaction ranges $L_f$ and $L_c$ are strongly dependent on improved understanding on how above-ground architectural units connect to the different components of the patch rhizosphere. On the dynamical side, it is necessary to carry out experiments as to verify how isolated small bunches of ramets react to variation in soil water resource and identify which level of soil moisture corresponds to the critical value of $\mu=1$. It is at this stage really interesting to have a reference model and well-defined theoretical predictions to guide field data collection in the future. It is all the more interesting that vegetation patterns are known to react to changes in both climate and anthropic influences ([15],[16],[23],[73],[74],[3]), and notably systems of localized structures may be thought of as the last stage before a possible collapse of vegetation cover and associated ecological functions and services.


\end{document}